\documentclass[aps,prl,preprint,groupedaddress]{revtex4-2}
\usepackage{latexsym}
\usepackage{graphicx}
\usepackage{float}

\begin{document}
\pagestyle{myheadings}


\title{Evolution of Octupole Deformation in Radium Nuclei from Coulomb Excitation of Radioactive $^{222}$Ra and $^{228}$Ra Beams}



\author{P.A.~Butler$^1$}
\email{peter.butler@liverpool.ac.uk}

\author{L.P.~Gaffney$^{1,2}$}

\author{P.~Spagnoletti$^3$}

\author{K.~Abrahams$^4$}

\author{M.~Bowry$^{3,5}$}

\author{J~.Cederk{\"a}ll$^6$}

\author{G.~de Angelis$^7$}

\author{H.~De~Witte$^8$}

\author{P.E.~Garrett$^9$}

\author{A.~Goldkuhle$^{10}$}

\author{C.~Henrich$^{11}$}

\author{A.~Illana$^7$}

\author{K.~Johnston$^2$}

\author{D.T.~Joss$^1$}

\author{J.M.~Keatings$^3$}

\author{N.A.~Kelly$^3$}

\author{M.~Komorowska$^{12}$}

\author{J.~Konki$^2$}

\author{T.~Kr{\"o}ll$^{11}$}

\author{M.~Lozano$^2$}

\author{B.S.~Nara Singh$^3$}

\author{D.~O\textsc{\char13}Donnell$^3$}

\author{J.~Ojala$^{13,14}$}

\author{R.D.~Page$^1$}

\author{L.G.~Pedersen$^{15}$}

\author{C.~Raison$^{16}$}

\author{P.~Reiter$^{10}$}

\author{J.A.~Rodriguez$^2$}

\author{D.~Rosiak$^{10}$}

\author{S.~Rothe$^2$}

\author{M.~Scheck$^3$}

\author{M.~Seidlitz$^{10}$}

\author{T.M.~Shneidman$^{17}$}

\author{B.~Siebeck$^{10}$}

\author{J.~Sinclair$^3$}

\author{J.F.~Smith$^3$}

\author{M.~Stryjczyk$^8$}

\author{P.~Van Duppen$^8$}

\author{S.~Vinals$^{18}$}

\author{V.~Virtanen$^{13,14}$}

\author{N.~Warr$^{10}$}

\author{K.~Wrzosek-Lipska$^{12}$}

\author{M. Zieli{\'n}ska$^{19}$}

\affiliation{
$^1$\mbox{University of Liverpool, Liverpool L69 7ZE, United Kingdom}\\
$^2$\mbox{ISOLDE, CERN,  1211 Geneva 23, Switzerland}\\
$^3$\mbox{University of the West of Scotland, Paisley PA1 2BE, United Kingdom}\\
$^4$\mbox{University of the Western Cape, Private Bag X17, Bellville 7535, South Africa}\\
$^5$\mbox{TRIUMF, Vancouver V6T 2A3 BC, Canada}\\
$^6$\mbox{Lund University, Box 118, Lund SE-221 00, Sweden}\\
$^7$\mbox{INFN Laboratori Nazionali di Legnaro, Legnaro 35020 PD, Italy}\\
$^8$\mbox{KU Leuven, Leuven B-3001, Belgium}\\
$^9$\mbox{University of Guelph, Guelph N1G 2W1 Ontario, Canada}\\
$^{10}$\mbox{University of Cologne, Cologne 50937, Germany}\\
$^{11}$\mbox{Technische Universit{\"a}t Darmstadt, Darmstadt 64289, Germany}\\
$^{12}$\mbox{Heavy Ion Laboratory, University of Warsaw, Warsaw PL-02-093, Poland}\\
$^{13}$\mbox{University of Jyvaskyla, P.O. Box 35, Jyvaskyla FIN-40014, Finland}\\
$^{14}$\mbox{Helsinki Institute of Physics, P.O. Box 64, Helsinki, FIN-00014, Finland}\\
$^{15}$\mbox{University of Oslo, P.O. Box 1048, Oslo N-0316, Norway}\\
$^{16}$\mbox{University of York, York YO10 5DD, United Kingdom}\\
$^{17}$\mbox{Joint Institute for Nuclear Research, RU-141980 Dubna, Russian Federation}\\
$^{18}$\mbox{Consejo Superior De Investigaciones Cient{\'i}ficas, Madrid S28040, Spain}\\
$^{19}$\mbox{IRFU CEA, Universit{\'e} Paris-Saclay, Gif-sur-Yvette F-91191, France}\\
}

\begin{abstract}
There is sparse direct experimental evidence that atomic nuclei can
exhibit stable `pear' shapes arising from strong octupole
correlations. In order to investigate the nature of octupole
collectivity in radium isotopes, electric octupole ($E$3) matrix
elements have been determined for transitions in $^{222,228}$Ra
nuclei using the method of sub-barrier, multi-step Coulomb
excitation. Beams of the radioactive radium isotopes were provided
by the HIE-ISOLDE facility at CERN.  The observed pattern of $E$3
matrix elements for different nuclear transitions is explained by
describing $^{222}$Ra as pear-shaped  with stable octupole
deformation, while $^{228}$Ra behaves like an octupole vibrator.
\end{abstract}

\maketitle

There are many theoretical and experimental indications that atomic
nuclei can exhibit reflection asymmetry in the intrinsic frame, and
observation of low-lying quantum states in many nuclei with even Z,
N having total angular momentum and parity of $I^\pi=3^-$ is
indicative of the presence of octupole correlations
(see~\cite{butl96} and references therein). Typically, the electric
octupole ($E$3) moment for the transition to the ground state is
tens of single-particle units, suggesting that the octupole
instability arises from a collective effect and leads to a
pear-shaped distortion of the nuclear shape. What is less clear,
however, is whether in some nuclei this distortion is stable, i.e.
the nucleus assumes a permanent pear shape, or whether it is dynamic
and the nucleus undergoes octupole vibrations.  Evidence has been
presented that $^{224}$Ra and $^{226}$Ra have static octupole
deformation on account of an enhancement in the $E$3 moment in these
nuclei~\cite{gaff13,woll93}. Large $E$3 moments have also been
recently measured for neutron-rich barium isotopes, suggesting that,
within the experimental uncertainty, these nuclei could have
octupole deformation~\cite{buch16,buch17}. The only example of an
octupole unstable nucleus other than $^{226}$Ra where stable beams
have been used to obtain a complete set of $E$3 matrix elements is
$^{148}$Nd~\cite{ibbo97}.

In this Letter, results from a multistep, Coulomb-excitation
experiment with radioactive $^{222, 228}$Ra beams are reported. By
examining the pattern of $E$3 matrix elements between different
transitions in these nuclei and comparing them to those in $^{224,
226}$Ra and $^{148}$Nd, a distinction can be made between those
isotopes having stable octupole deformation and those behaving like
octupole vibrators. This observation is relevant for the search for
permanent electric dipole moments in radium
atoms~\cite{auer96,doba05,bish16}, that would indicate sizeable CP
violation requiring a substantial revision of the Standard Model.

\begin{figure}
\includegraphics[width=0.8\columnwidth]{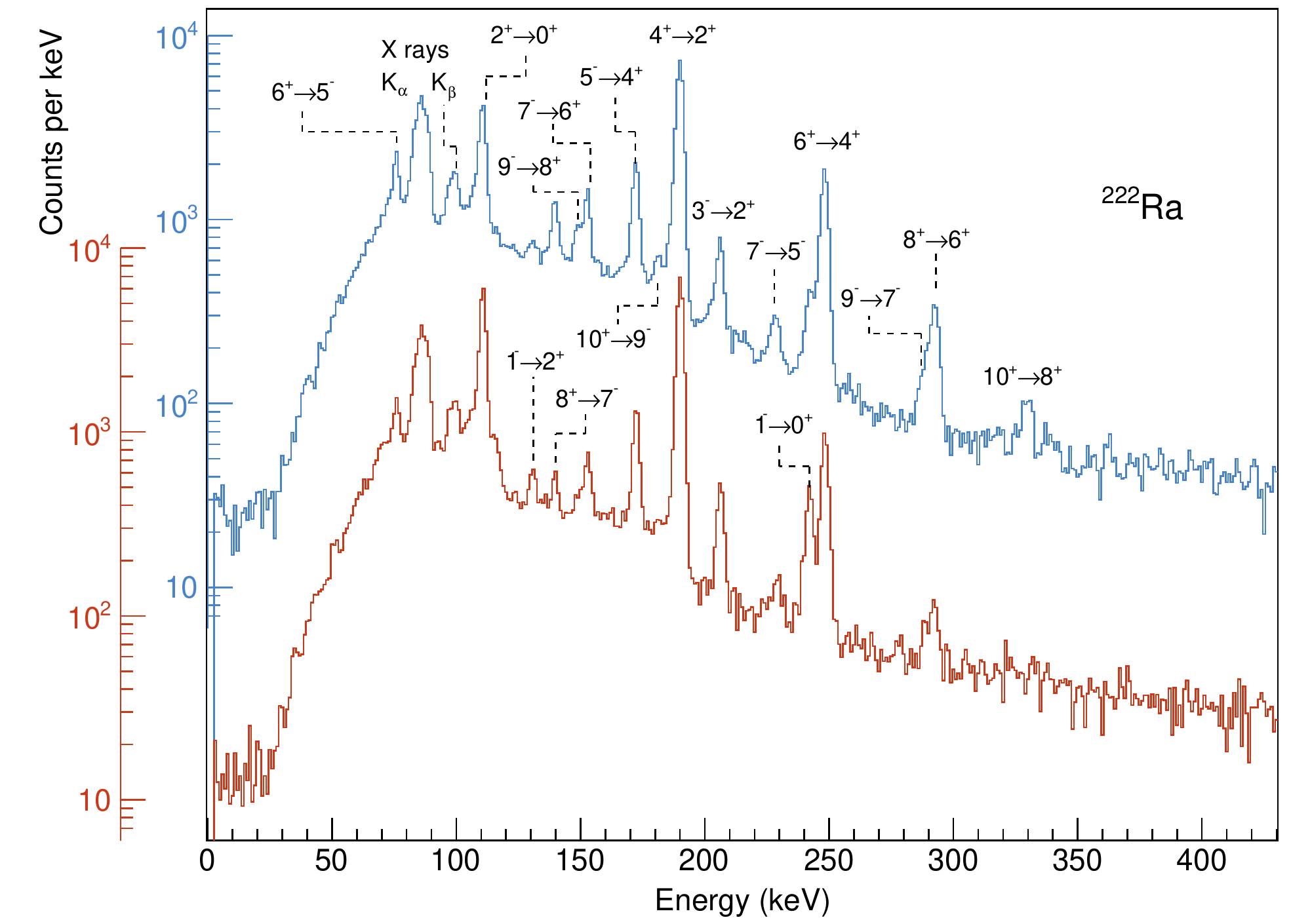}
\includegraphics[width=0.8\columnwidth]{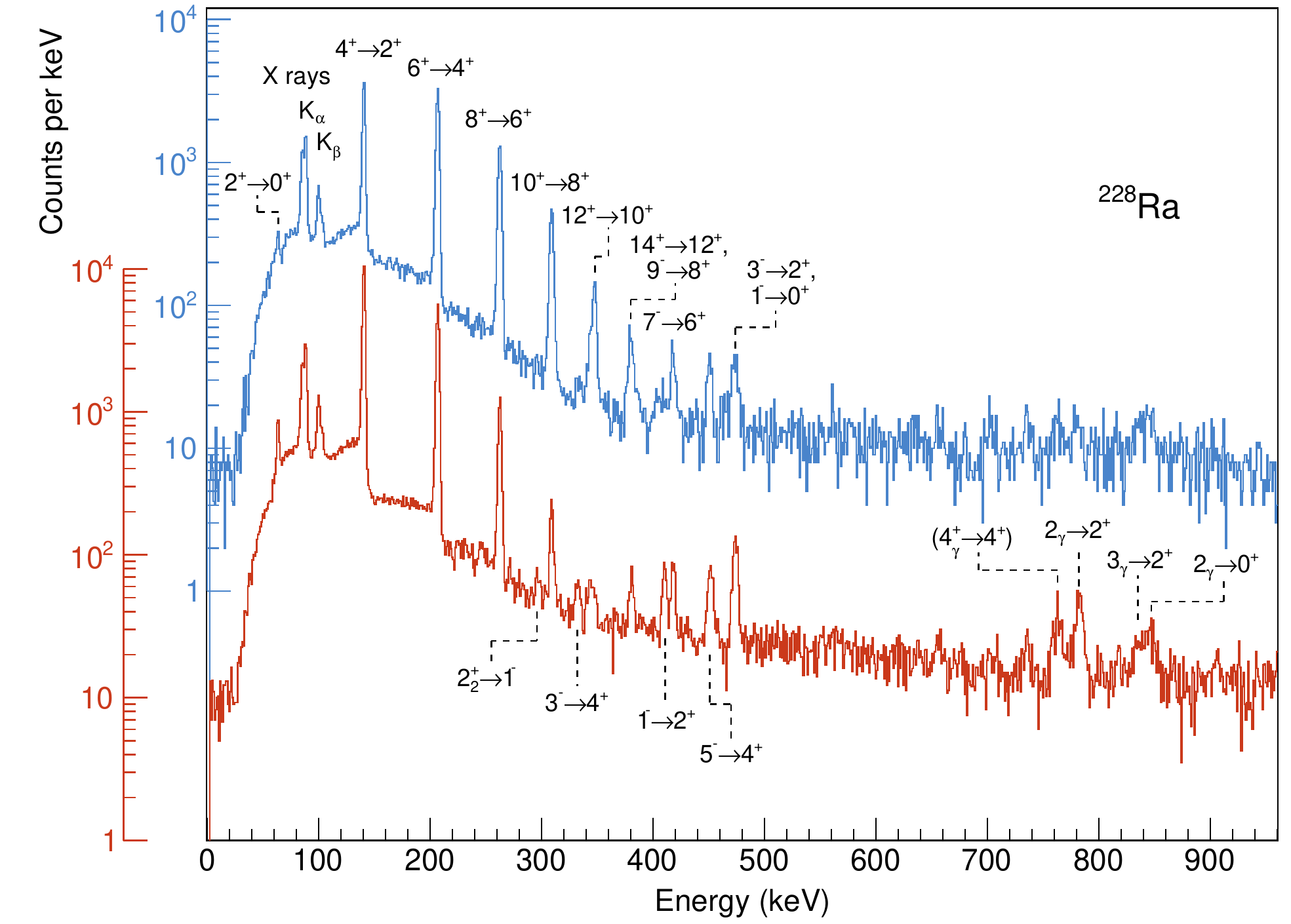}
\caption{\label{spectra}Spectra of $\gamma$ rays emitted following
the Coulomb excitation of $^{222}$Ra (upper) and $^{228}$Ra (lower)
using a $^{120}$Sn target (blue), and $^{60}$Ni (red). The $\gamma$
rays were corrected for Doppler shift assuming that they are emitted
from the scattered projectile. Random coincidences between Miniball
and the silicon detector have been subtracted. The transitions that
give rise to the observed full-energy peaks are labelled by the spin
and parity of the initial and final states.}
\end{figure}

The radioactive isotopes $^{222}$Ra (Z = 88, N = 134) and $^{228}$Ra
(Z = 88, N = 140) were produced by spallation in a thick uranium
carbide primary target bombarded by $\approx 10^{13}$ protons/s at
1.4 GeV from the CERN PS Booster. The ions, extracted from a
tungsten surface ion source were stripped to charge states of 51$^+$
and 53$^+$, respectively, for $^{222}$Ra and $^{228}$Ra and
accelerated in HIE-ISOLDE to an energy of 4.31 MeV/nucleon. The
radioactive beams, with intensities between $5\times 10^4$ and
$2\times 10^5$ ions/s bombarded secondary targets of $^{60}$Ni and
$^{120}$Sn of thickness 2.1 mg/cm$^2$.    Gamma rays emitted
following the excitation of the target and projectile nuclei were
detected in Miniball~\cite{warr13}, an array of 24 high-purity
germanium detectors, each with sixfold segmentation and arranged in
eight triple clusters. The scattered projectiles and target recoils
were detected in a highly segmented silicon detector, distinguished
by their differing dependence of energy with angle
measured in the laboratory frame of reference.  
Representative spectra from the Coulomb-excited $^{222,228}$Ra are
shown in Fig.~\ref{spectra}; in the spectra the $\gamma$-ray
energies are corrected for Doppler shift assuming emission from the
scattered projectile. The spectra were incremented when a target
recoil was detected in coincidence with $\gamma$ rays within a
450-ns time window; these data were corrected for random events. The
fraction of the isobar $^{222}$Fr in the beam was estimated to be
about 20\% by observing $\gamma$ rays from the $\alpha$-decay
daughters at the beam dump. By lowering the temperature of the
transfer line from the ion source a nearly pure beam of $^{222}$Fr
could be produced; apart from X-rays, no discernable structure was
observed arising from Coulomb excitation of the odd-odd nucleus in
the particle-gated, Doppler-corrected spectrum. For the $^{228}$Ra
beam, the fraction of isobaric contamination was estimated to be
$\approx 1\%$.

For both $^{222}$Ra and $^{228}$Ra the spectra reveal strong
population of the ground-state band of positive-parity states,
populated by multiple electric quadrupole ($E$2) Coulomb excitation,
and substantial population of the octupole band of negative-parity
states, populated by $E$3 excitation. The yields of the observed
$\gamma$-ray transitions detected in Miniball were measured for four
ranges of the recoil angle of the target nucleus for each target,
between $21.5^{\circ}$ and $55.5^{\circ}$ for the $^{120}$Sn target
and between $17.8^{\circ}$ and $55.5^{\circ}$ for the $^{60}$Ni
target. The yield data were combined with existing $\gamma$-ray
branching ratios to provide input to the Coulomb-excitation analysis
code GOSIA~\cite{czos83,clin93,ziel16}. The GOSIA code performs a
least-squares fit to the $E\lambda$ ($\lambda =1,2,3$) matrix
elements (m.e.s), which either can be treated as free parameters,
can be coupled to other matrix elements, or can be fixed.
Energy-level schemes that are included in the analysis are given
in~\cite{supp}. A total of 114 data for $^{222}$Ra were fitted to 42
variables, while for $^{228}$Ra 121 data were fitted to 41
variables. The starting values of each of the freely-varied matrix
elements were drawn randomly, within reasonable limits; the values
obtained following the fitting procedure were found to be
independent of the starting points. Examples of fits to the
experimental data can be found in the Supplemental Material, see
below~\cite{supp}.

For both nuclei the $E$1 couplings between the ground-state and
negative-parity bands and the $E$2 couplings for transitions within
the ground state and within the negative-parity bands, with the
exception of the $2^+ \rightarrow 0^+$ transition, were treated as
free parameters. Under the experimental conditions described here,
the probability of populating the $2^+$ state is $> 90 \%$ and it
was not possible to determine the $\langle 0^{+}||E2||2^{+}\rangle$
and $\langle 2^{+}||E2||2^{+}\rangle$ m.e.s independently. The
latter was therefore allowed to vary freely and the $\langle
0^{+}||E2||2^{+}\rangle$ matrix element was coupled to the $\langle
2^{+}||E2||4^{+}\rangle$ matrix element assuming the validity of the
rotational model; this assumption is based on the behaviour of
nuclei where the lifetimes of the $2^+$ and $4^+$ states have been
measured and for which the lowest transitions behave
collectively~\cite{supp}. For the $E$3 m.e.s the lowest couplings
were treated as free parameters; m.e.s between higher-lying states,
$\langle I^{\pm}||E3||I'^{\mp}\rangle$, were coupled to m.e.s
between lower-lying states,
$\langle(I-2)^{\pm}||E3||(I'-2)^{\mp}\rangle$, assuming the validity
of the rotational model. $E$4 matrix elements were also included in
the fitting procedure; these were calculated assuming the rotational
model and a constant value of the hexadecapole moment, derived from
the theoretical values of $\beta_\lambda$~\cite{naza84}. $E$2 (and
magnetic dipole) couplings to high-lying $K^{\pi} = 0^+$  and
$K^{\pi} = 2^+$ bands were also taken into account.   The relative
phase of $\mathcal{Q}_1$ and $\mathcal{Q}_3$ was investigated, as
although the overall phase of the $E$1 and $E$3 matrix elements is
arbitrary, the fit is sensitive to the relative phase of $E$3 matrix
elements as well as the phase difference between the $E$1 and $E$3
matrix elements. The difference in chi-square for the fit favoured
$\mathcal{Q}_1$ and $\mathcal{Q}_3$ having the same sign for
$^{222}$Ra and the opposite sign for $^{228}$Ra, and these phases
were adopted in the final fits. These values are consistent with
macroscopic-microscopic calculations~\cite{butl91} and constrained
HFBCS calculations~\cite{egid89} that predict a decreasing value of
$\mathcal{Q}_1$ with neutron number for radium isotopes, crossing
zero for $^{224}$Ra as experimentally verified~\cite{poyn89}.

Table~\ref{results_E2_E3} gives the values of $E$2 and $E$3 matrix
elements for $^{222}$Ra and $^{228}$Ra obtained in this work. The
$E$1 matrix elements are given in~\cite{supp}. Those $E$3 m.e.s
marked with an asterisk are coupled to m.e.s between higher-lying
states and as such are not completely independently determined;
however the fit is mostly influenced by the value of the lowest
matrix element. The diagonal $E$2 matrix elements are all coupled to
the adjacent transition m.e.s except for those presented in
Table~\ref{results_E2_E3}, which are independently determined. In
the GOSIA fit the statistical errors for each fitted variable were
calculated taking into account correlations between all variables.
Independent sets of fitted values were also obtained by varying the
constant hexadecapole moment used to calculate the $E$4 m.e.s
between zero and double the notional value, varying the target
thickness by $\pm 5\%$, the beam energy by $\pm 1\%$ , the distance
between the target and the particle detector by $\pm 7.5\%$, and the
sign of the $E$2 couplings to the higher-lying collective bands.

\begin{figure}
\includegraphics[width=0.8\columnwidth]{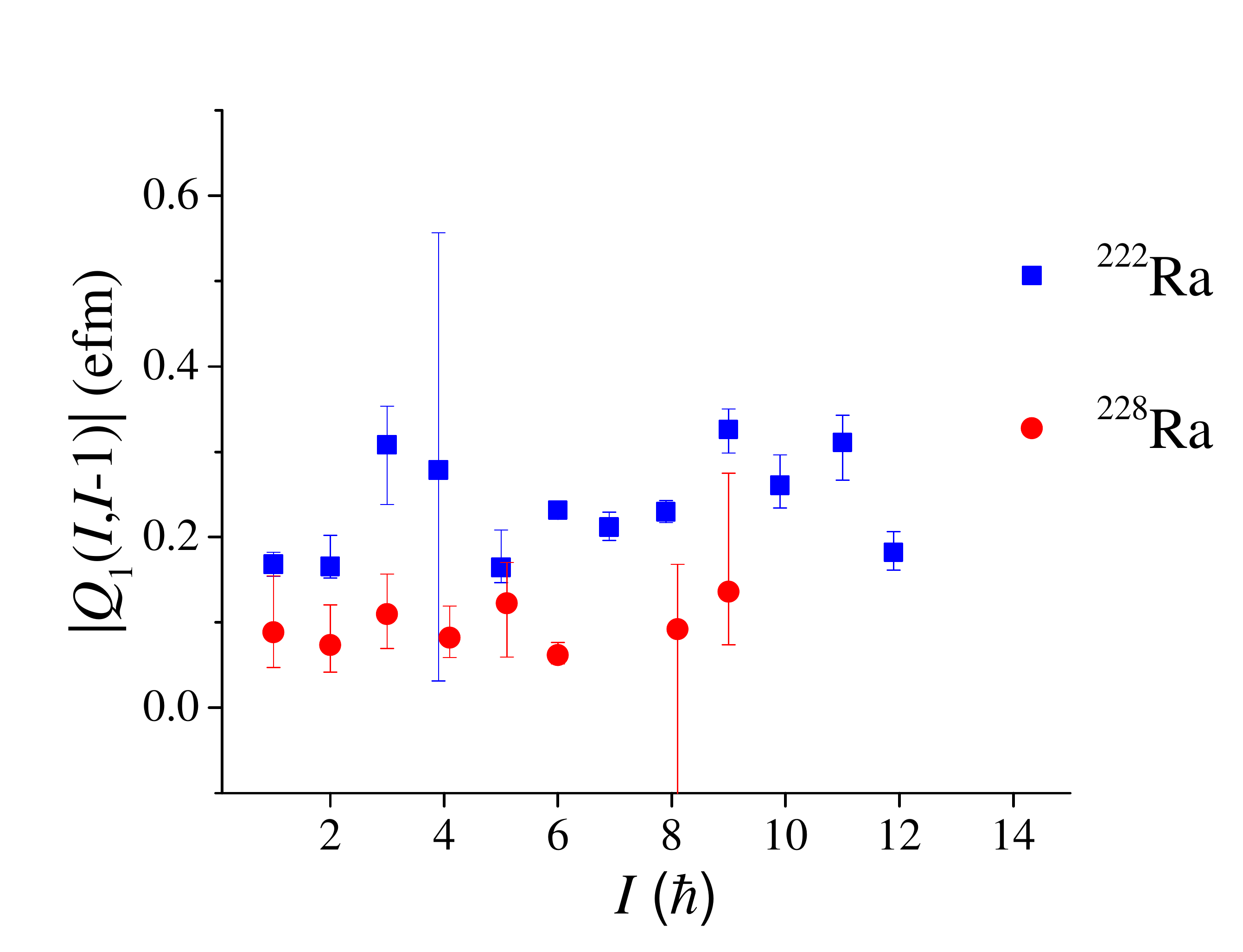}
\caption{\label{Q_1} Absolute values of the intrinsic dipole
moments, $\mathcal{Q}_1$ as a function of spin. The values are
deduced from the measured matrix elements~\cite{supp}, and
correspond to transitions between states with spin $I$ and $I-1$. }
\end{figure}
\begin{figure}
\includegraphics[width=0.8\columnwidth]{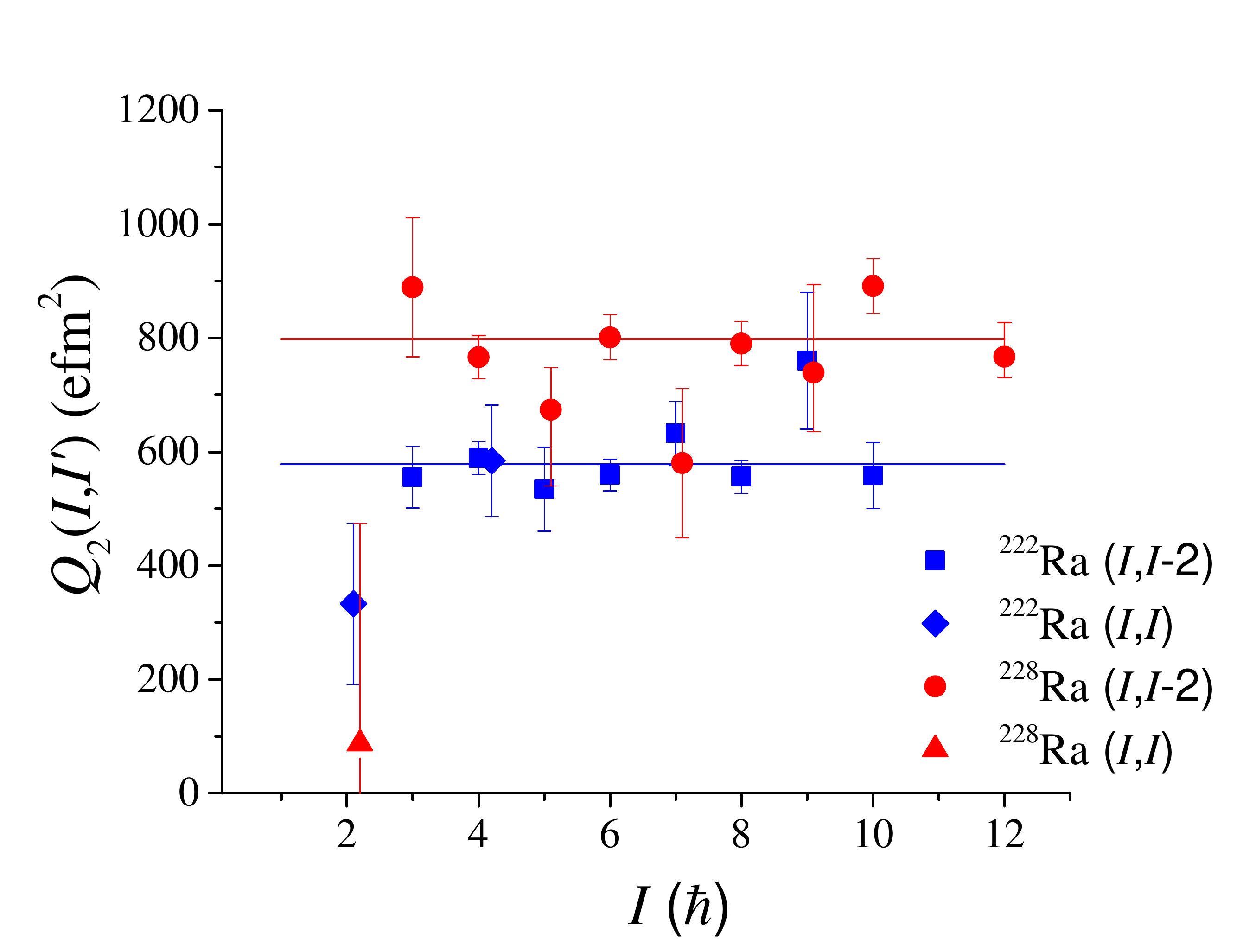}
\caption{\label{Q_2} Values of the intrinsic quadrupole moments,
$\mathcal{Q}_2$ plotted as a function of spin. The values are
deduced from the measured matrix elements given in
Table~\ref{results_E2_E3}. The values correspond to transitions
between states with spin $I$ and $I-2$; in some cases they are also
derived from diagonal matrix elements. The solid horizontal lines
correspond to the values of $\mathcal{Q}_2$ obtained assuming that
the matrix elements are related by the rotational model. }
\end{figure}

The variations seen in the fitted values are included in the final
uncertainties given in Table~\ref{results_E2_E3}. For $^{228}$Ra the
value of the intrinsic quadrupole moment, $\mathcal{Q}_2$, derived
from the measured value of $\langle 2^{+}||E2||4^{+}\rangle$, $770
\pm 40$ efm$^2$, agrees with the values determined from the $2^+$
lifetime, $775 \pm 14$ efm$^2$ and the $4^+$ lifetime, $780 \pm 6$
efm$^2$, as reported in Ref.~\cite{abus14}.
 For $^{222}$Ra, the value is $590 \pm 30$ efm$^2$, significantly smaller than the
value derived from the measured lifetime of the $2^+$ state, $673
\pm 13$ efm$^2$~\cite{bell60}. It is noted that the value of
$\mathcal{Q}_2$ for $^{222}$Ra extrapolated from the $2^+$ lifetime
for  $^{228}$Ra on the basis of B($E2; 0^+ \rightarrow 2^+$)
systematics~\cite{prit17}, is $593 \pm 11$ efm$^2$, in good
agreement with the current measurement. Fitted values of
$\mathcal{Q}_2$ and $\mathcal{Q}_3$ assuming that the $E\lambda$
matrix elements and $\mathcal{Q}_\lambda$ are related by the
rotational model are also given in Table~\ref{results_E2_E3}. The
values for $\lambda =3$ indicate that the octupole collectivity in
$^{228}$Ra is significantly lower than for $^{222}$Ra.
 \begin{table*}
 \caption{\label{results_E2_E3} Values of $E$2 and $E$3 matrix elements measured in the present experiment.
 The intrinsic moments, $\mathcal{Q}_\lambda$, are derived from each matrix element (m.e.) using
 $\langle I_i || \mbox{$\cal{M}$} (E{\lambda}) || I_f \rangle = \sqrt{(2I_i
+1)} \sqrt{(2\lambda+1)/16\pi} (I_i 0 \lambda 0 | I_f 0)
Q_{\lambda}$.
 The uncertainties include the $1\sigma$  statistical error from the fit ($\chi^2 + 1$ type) and the
 systematic contributions. The $E$3 m.e.s
marked with an asterisk are coupled to higher-lying m.e.s.
 The $\langle 0^{+}||E2||2^{+}\rangle$ and $\langle 2^{+}||E2||4^{+}\rangle$ m.e.s are coupled. Values of
$\mathcal{Q}_\lambda$ fitted assuming that the m.e.s are related by
the rotational model are also given.}
\begin{ruledtabular}
\begin{tabular}{ccccc}

& $^{222}$Ra & &  $^{228}$Ra & \\
$\langle I ||E\lambda|| I'\rangle$ & m.e. & $\mathcal{Q}_\lambda$ & m.e.   & $\mathcal{Q}_\lambda$ \\
&  (eb$^{\lambda/2}$) &  (efm$^{\lambda}$) & (eb$^{\lambda/2}$) &  (efm$^{\lambda}$) \\
\hline
$\langle 2^{+}||E2||2^{+}\rangle$ & $-1.3 \pm 0.5$ & $330 \pm 140$ & $-0.3 \pm 1.7$ & $90 \pm 400$\\
$\langle 2^{+}||E2||4^{+}\rangle$ & $2.98 \pm 0.15$   & $590 \pm 30$ & $3.87 \pm   0.19$ & $770 \pm 40$\\
$\langle 4^{+}||E2||4^{+}\rangle$ & $-2.8 \pm 0.5$  & $580 \pm 100$ &  &  \\
$\langle 4^{+}||E2||6^{+}\rangle$ & $3.57 \pm 0.18$  & $559 \pm 28$ & $5.11 \pm 0.26$ & $800 \pm 40$ \\
$\langle 6^{+}||E2||8^{+}\rangle$ & $4.15 \pm 0.23$ & $560 \pm 30$ & $5.89 \pm 0.29$ & $790 \pm 40$ \\
$\langle 8^{+}||E2||10^{+}\rangle$ & $4.7 \pm 0.5$ & $560 \pm 60$ & $7.5 \pm 0.4$ &  $890 \pm 50$ \\
$\langle 10^{+}||E2||12^{+}\rangle$ &   & & $7.1^{+0.5}_{-0.3}$ & $770 ^{+60}_{-40}$ \\
$\langle 1^{-}||E2||3^{-}\rangle$ & $2.35 \pm  0.22$  & $560 \pm 50$& $3.8 \pm   0.5$ & $890 \pm 120$\\
$\langle 3^{-}||E2||5^{-}\rangle$ & $3.1 \pm 0.4$  & $530 \pm 70$ & $3.9 ^{+0.4}_{-0.8}$ & $670 ^{+70}_{-130}$\\
$\langle 5^{-}||E2||7^{-}\rangle$ & $4.4 \pm  0.4$   & $630 \pm 60$ & $4.0 \pm   0.9$ & $580 \pm 130$\\
$\langle 7^{-}||E2||9^{-}\rangle$ &  $6.0 \pm 1.0$  & $760 \pm 120$ & $5.9 \pm   1.0$ & $740 \pm 130$\\
$\mathcal{Q}_2$ (rotational model)&  &  $578 \pm 18$ &  &  $798 \pm 21$\\
 \hline
$\langle 0^{+}||E3||3^{-}\rangle$ & $1.13 \pm 0.09$ & $3030 \pm 240$  & $0.87 \pm 0.15$ & $2300 \pm 400$\\
$\langle 2^{+}||E3||1^{-}\rangle$ & $0.85 \pm 0.24$ & $2000 \pm 600$  & $1.36 \pm 0.23$* & $3200 \pm 600$\\
$\langle 2^{+}||E3||3^{-}\rangle$ & $-0.9 \pm 0.5$ & $2100 \pm 1200$  & $-0.06^{+0.23}_{-0.16}$* & $ 150 ^{+360}_{-500}$\\
$\langle 2^{+}||E3||5^{-}\rangle$ & $1.79 \pm 0.20$ & $3100 \pm 400$  & $1.71 \pm 0.23$* & $3000 \pm 400$\\
$\langle 4^{+}||E3||1^{-}\rangle$ & $-2.1 \pm 0.5$* & $4400 \pm 1000$  & $0.4^{+0.7}_{-1.1}$* & $ -800^{+2300}_{-1400}$\\
$\langle 4^{+}||E3||3^{-}\rangle$ & $2.6 ^{+0.6}_{-0.9}$*  &  $5500 ^{+1300}_{-1800}$ &  &\\
$\langle 4^{+}||E3||5^{-}\rangle$ & $-1.7 \pm 1.0$* & $3200 \pm 1800$ &  & \\
$\langle 4^{+}||E3||7^{-}\rangle$ & $3.3 ^{+0.3}_{-0.5}$* & $4600^{+500}_{-600}$ &  & \\
$\mathcal{Q}_3$ (rotational model) &  &  $3120 \pm 190$ &  &  $2230 \pm 290$\\

 \end{tabular}

 \end{ruledtabular}
 \end{table*}

\begin{figure}[h]
\includegraphics[width=0.8\columnwidth]{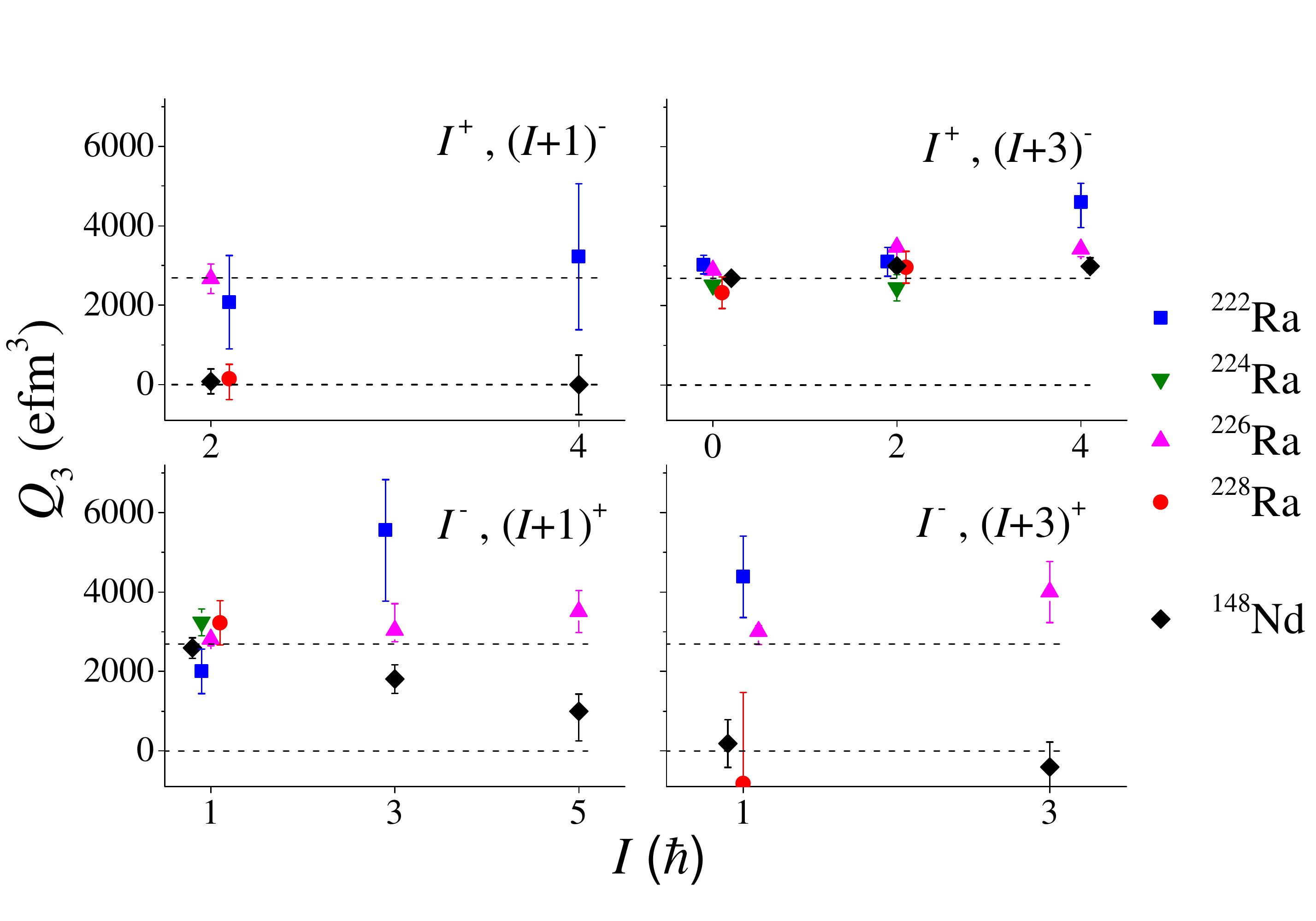} \caption{\label{Q3}
Values of the intrinsic octupole moments, $\mathcal{Q}_3$
plotted as a function of spin. The values are deduced from the
measured matrix elements given in Table~\ref{results_E2_E3}. Here
the values of $\mathcal{Q}_3$ are shown separately for transitions
connecting $I^{+} \rightarrow (I+1)^{-}$, $I^{+} \rightarrow
(I+3)^{-}$, $I^{-}\rightarrow (I+1)^{+}$ and $I^{-}\rightarrow
(I+3)^{+}$. The upper dashed line is the average value of
$\mathcal{Q}_3 (0^+ , 3^-)$ for the radium isotopes. To aid
comparison the values of $\mathcal{Q}_3$ for $^{148}$Nd have been
multiplied by 1.78. }
\end{figure}

The values of $\mathcal{Q}_1$ and $\mathcal{Q}_2$ for all the
measured matrix elements are shown in Figs.~\ref{Q_1} and \ref{Q_2},
respectively. The nearly constant values of $\mathcal{Q}_2$ as a
function of spin for transitions in both positive- and
negative-parity bands is consistent with stable quadrupole
deformation. Smaller values of $\mathcal{Q}_2$, although with large
uncertainty, were determined from the $\langle
2^{+}||E2||2^{+}\rangle$ matrix element for both nuclei. Such
behaviour was also observed in $^{226}$Ra, interpreted as arising
from deviations from axial symmetry~\cite{woll93}. The values of the
intrinsic electric octupole moment $\mathcal{Q}_3$ for transitions
in $^{222}$Ra and $^{228}$Ra are shown in Fig.~\ref{Q3}.
In the figure, the values of $\mathcal{Q}_3$ are shown separately
for transitions  $I^{+} \rightarrow (I+1)^{-}$, $I^{+} \rightarrow
(I+3)^{-}$, $I^{-}\rightarrow (I+1)^{+}$ and $I^{-}\rightarrow
(I+3)^{+}$, and are compared with values determined for the same
transitions in $^{224,226}$Ra~\cite{gaff13,woll93} and
$^{148}$Nd~\cite{ibbo97}. The values for $^{148}$Nd are multiplied
by a factor so that the value of $\mathcal{Q}_3$ deduced from
$\langle 0^{+}||E3||3^{-}\rangle$ is the same as the average value
for the radium isotopes. It is observed that the values of
$\mathcal{Q}_3$ for all transitions in $^{222,224,226}$Ra are
approximately constant, consistent with the picture of a rotating
pear shape. In contrast, the values of $\mathcal{Q}_3$ corresponding
to the $2^{+} \rightarrow 3^{-}$ and $1^{-} \rightarrow 4^{+}$
transitions in $^{228}$Ra are close to zero, as observed for
$^{148}$Nd. It is unlikely that this can be accounted for by $K$
mixing~\cite{clin93} as the $K^\pi=1^-$ band lies much higher in
energy for these nuclei~\cite{neer70}.

\begin{figure}
\includegraphics[width=0.8\columnwidth]{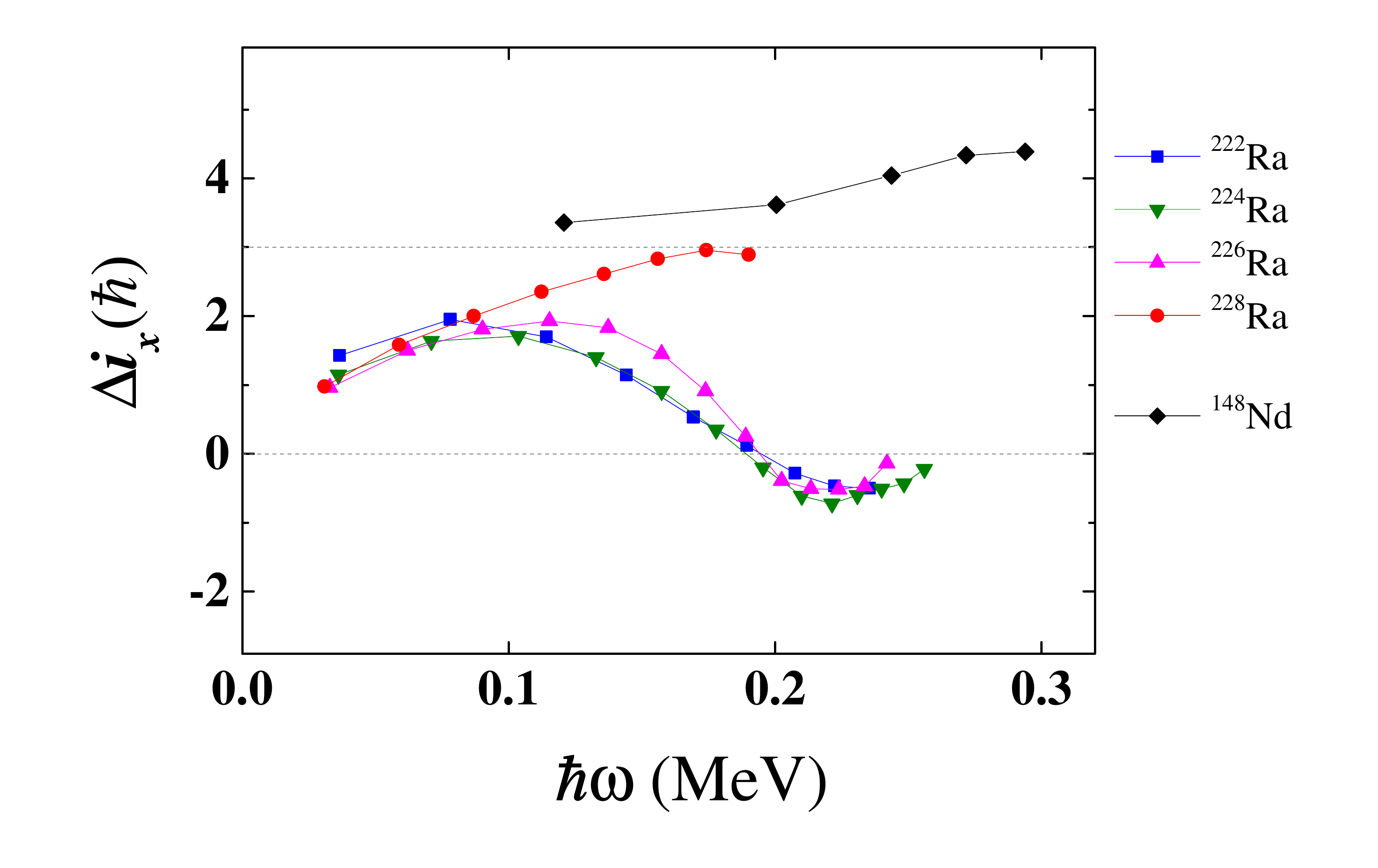}
\vspace{-0.75cm} \caption{\label{delta_i} The difference in aligned
angular momentum, $\Delta i_{x} = i_{x}^- \, - \, i_{x}^+ $, plotted
as a function of rotational frequency $\omega$. The upper dashed
line corresponds to the vibrational limit, $\Delta i_{x} = 3 \hbar$}
\end{figure}

The contrast in the behaviour of the $E$3 moments of $^{228}$Ra (and
$^{148}$Nd) compared to the lighter radium isotopes is also present
in the behaviour of their energy levels, as shown in
Fig.~\ref{delta_i}.
Here $\Delta i_{x}$, the difference in aligned angular momentum
between negative- and positive-parity states at the same rotational
frequency $\omega$, is plotted as a function of $\hbar \omega$ for
the five nuclei. The behaviour of $\Delta i_{x}$ can reveal
information regarding the nature of the octupole
correlations~\cite{cock97,butl16}. For $^{148}$Nd, the value of
$\Delta i_{x} \sim 3 \hbar$  for all values of rotational frequency,
and for $^{228}$Ra it approaches $3 \hbar$ when $\hbar \omega
\rightarrow 0.15$~MeV. This behaviour is expected for octupole
vibrators, where the octupole phonon aligns to the rotation axis. It
is conjectured here that the observation of near-zero values of
$\mathcal{Q}_3$ for some transitions in $^{228}$Ra (and $^{148}$Nd)
is consistent with the octupole-vibrator description. The
interpretation of the behaviour of energy levels for
$^{222,224,226}$Ra in terms of rotating pear shapes is less obvious
as it is dominated by pairing effects near the ground state; other
interpretations of this behaviour, e.g. the condensation of
rotational-aligned octupole phonons~\cite{frau08}, do not require
the nucleus to have a permanent octupole distortion. On the other
hand highly-collective $E$2 and $E$3 transition strengths are nearly
independent of pairing and single particle effects and are a much
better measure of the nuclear shape. The observed enhancement and
rotorlike pattern of the electric octupole moments $\mathcal{Q}_3$
provide compelling evidence that $^{222}$Ra  together with
$^{224,226}$Ra have stable octupole deformation. This confirms
theoretical predictions, e.g.~\cite{naza84,robl13,agbe16}, that the
boundary of octupole deformation lies at Z $\approx 88$ and at N
$\approx 138$; it has already been established that even-even radon
(Z=86) nuclei having similar neutron numbers behave like octupole
vibrators~\cite{butl19}. It is concluded that the differing patterns
of $E$3 matrix elements observed for $^{222,228}$Ra are a
consequence of the stability of the octupole shape for each nucleus.
Any model of quadrupole-octupole coupling that describes this
behaviour should be capable of calculating values of $\mathcal{Q}_3$
for different $E$3 transitions including the critical $3^-
\rightarrow 2^+$ transition, as has been performed for
$^{224}$Ra~\cite{xia17}.

\begin{acknowledgments}
We are grateful to Doug Cline and the late Tomek Czosnyka who led
the development of the Coulomb excitation analysis technique used in
this work, and to Niels Bidault, Eleftherios Fadakis, Erwin
Siesling, and Fredrick Wenander who assisted with the preparation of
the radioactive beams. The support of the ISOLDE Collaboration and
technical teams is acknowledged. This work was supported by the
following Research Councils and Grants: Science and Technology
Facilities Council (UK) Grants No. ST/P004598/1, No. ST/L005808/1,
No. ST/R004056/1; Federal Ministry of Education and Research
(Germany) Grants No. 05P18RDCIA, No. 05P15PKCIA, and No. 05P18PKCIA
and the ``Verbundprojekt 05P2018''; National Science Centre (Poland)
Grant No. 2015/18/M/ST2/00523; European Union's Horizon 2020
Framework research and innovation programme 654002 (ENSAR2); Marie
Sk{\l}odowska-Curie COFUND Grant (EU-CERN) 665779; Research
Foundation Flanders and IAP Belgian Science Policy Office BriX
network P7/12 (Belgium); GOA/2015/010 (BOF KU Leuven); RFBR (Russia)
Grant No. 17-52-12015; the Academy of Finland (Finland) Grant No.
307685.
\end{acknowledgments}

\clearpage

\section{Supplemental Material}

\begin{figure}[H]
\centerline{\includegraphics[width=0.8\columnwidth]{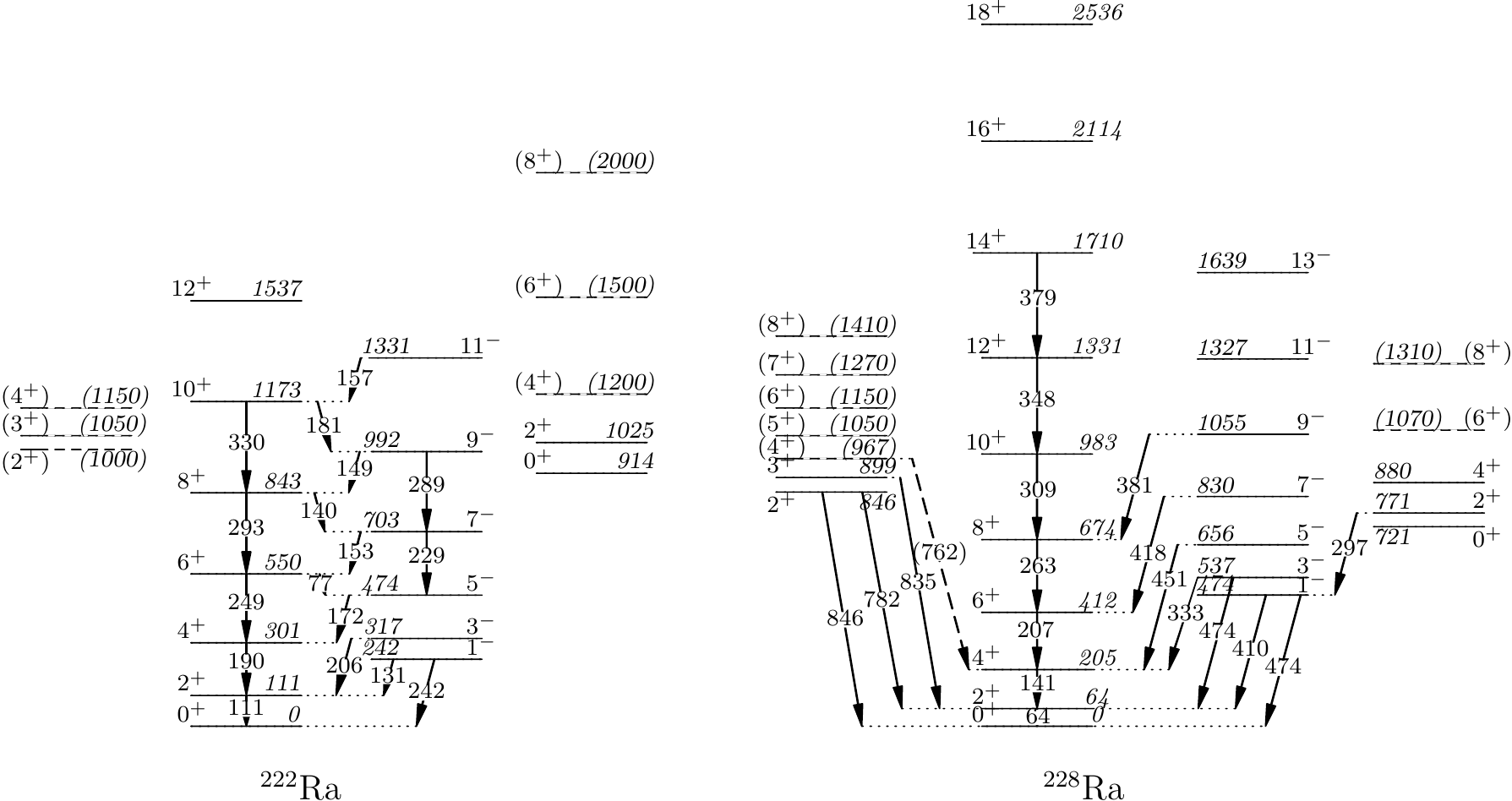}}
\caption{\label{level_schemes} Partial level-schemes for
$^{222,228}$Ra showing the excited states used in the
Coulomb-excitation analysis. The dashed levels, also included in the
fitting procedure, have either been tentatively labelled with spin
and parity or have been artificially constructed for this purpose.
Arrows indicate $\gamma$-ray transitions that have been observed in
the experiments described here; all energies are in keV. In
$^{222}$Ra no transitions to the higher lying collective bands were
observed. The level scheme data have been taken from
~\cite{sing11,abus14}. }
\end{figure}

Fig.~\ref{level_schemes} shows the partial level-schemes for
$^{222,228}$Ra used in the Coulomb-excitation analysis.

\newpage

\begin{figure}[H]
\includegraphics[width=0.45\columnwidth]{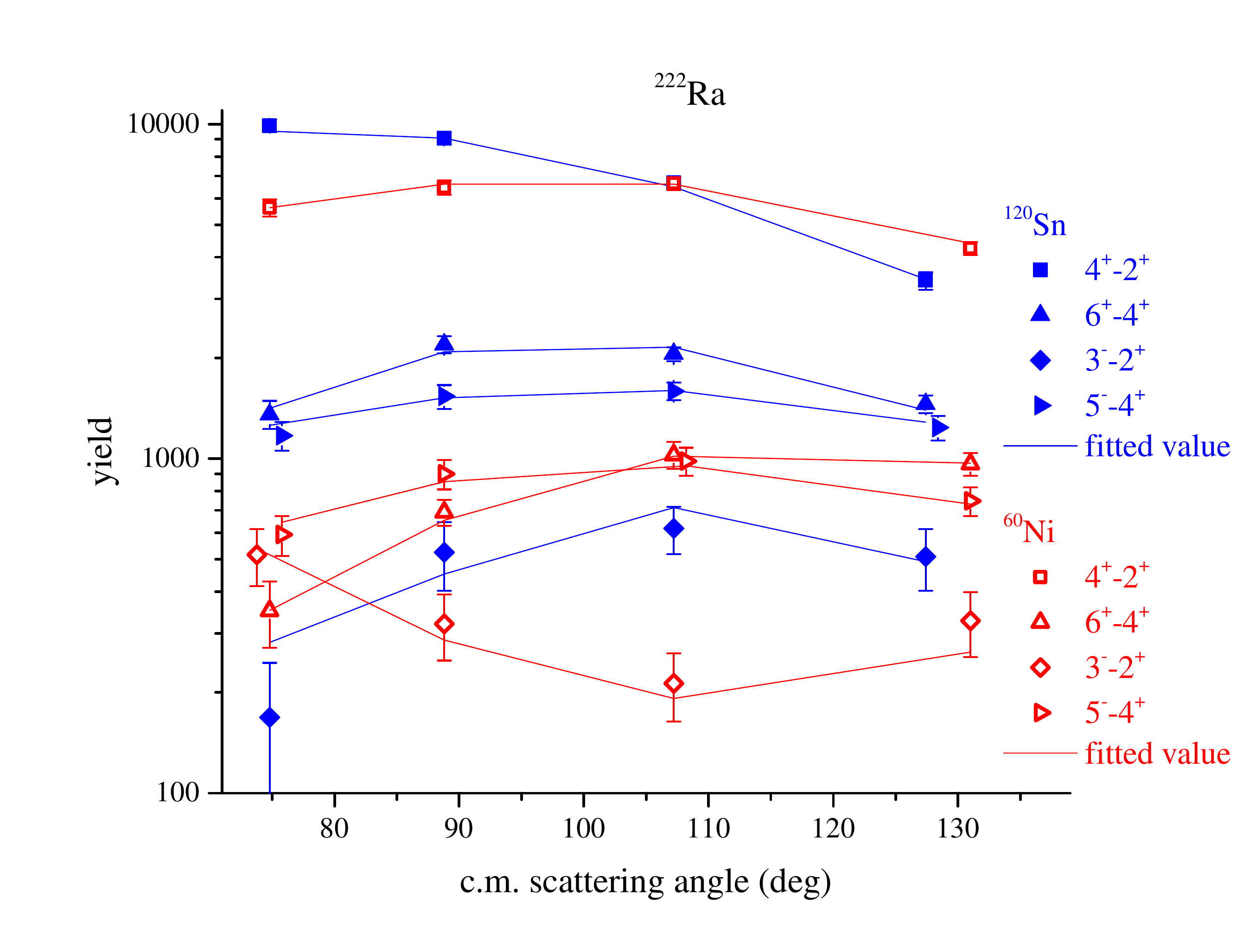}
\includegraphics[width=0.45\columnwidth]{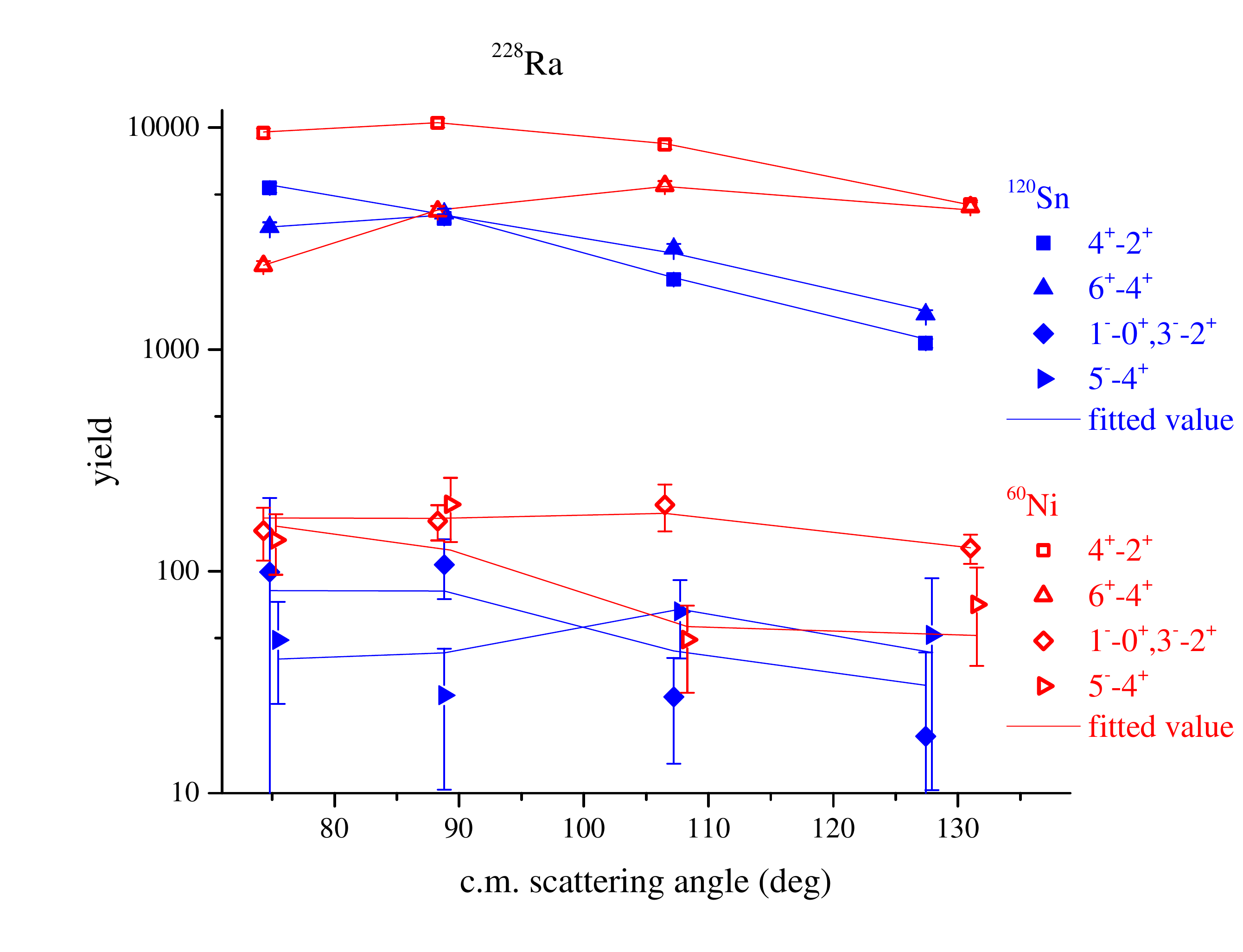}
\caption{\label{fits} Comparison of the experimental $\gamma$-ray
yields and uncertainties with those calculated with GOSIA for
selected transitions, based on the set of matrix elements resulting
in the best overall agreement with all the available experimental
data, including previously measured branching ratios. GOSIA takes
into account internal conversion using the {\it BrICC}
database~\cite{kibe08}.}
\end{figure}

Fig.~\ref{fits} shows a comparison of the experimental yields and
uncertainties for selected transitions with those calculated with
GOSIA based on the set of matrix elements resulting in the best
overall agreement with the experimental data.

\newpage

\begin{figure}[H]
\centerline{\includegraphics[width=0.8\columnwidth]{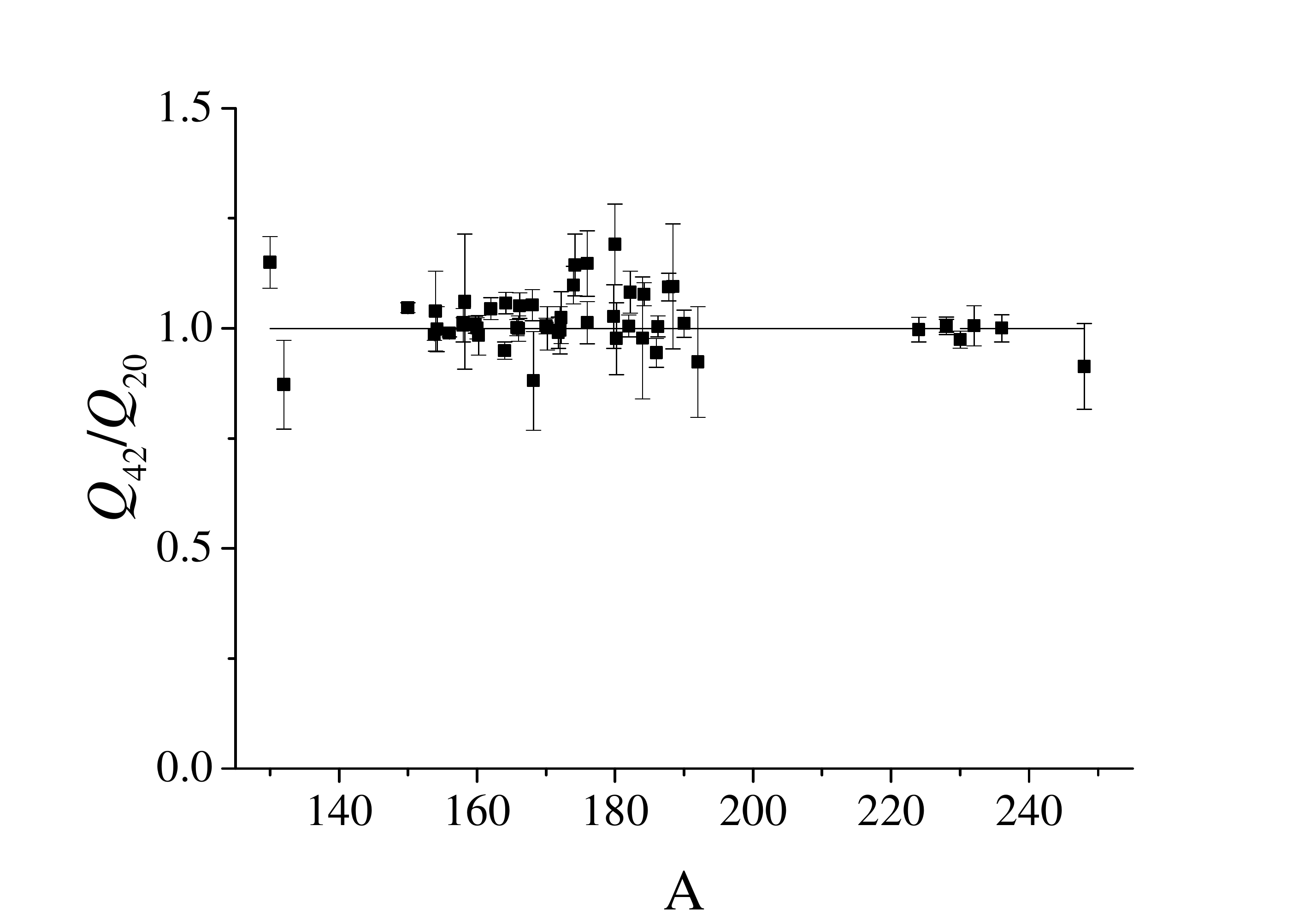}}
\caption{\label{Q42-Q20-ratio} Ratio of $\mathcal{Q}_{42}$ to
$\mathcal{Q}_{20}$ deduced from the transition strengths B($E2; 4^+
\rightarrow 2^+$) and B($E2; 2^+ \rightarrow 0^+$), respectively,
assuming the validity of the rotational model. This has been
determined for nuclei where the lifetimes of both $2^+$ and $4^+$
states have been measured. Only nuclei where $B(E2; 2^+ \rightarrow
0^+) > 70$~Wu and $B(E2; 4^+ \rightarrow 2^+) > 70$~Wu are included
in this compilation. The lifetime data have been taken
from~\cite{ensdf}. }
\end{figure}

Fig.~\ref{Q42-Q20-ratio} shows the ratio of $\mathcal{Q}_{42}$ to
$\mathcal{Q}_{20}$ deduced from the transition strengths B($E2; 4^+
\rightarrow 2^+$) and B($E2; 2^+ \rightarrow 0^+$), respectively,
assuming the validity of the rotational model, for nuclei with $A
> 130$ where the lifetimes of both $2^+$ and $4^+$ states have been
measured. Only nuclei where $B(E2; 2^+ \rightarrow 0^+) > 70$~Wu and
$B(E2; 4^+ \rightarrow 2^+) > 70$~Wu are included in this
compilation. The restriction to nuclei having collective low-energy
transitions does not reveal departures from rotational behaviour, as
observed in other studies, e.g.~\cite{caki04,thia06,sayg17}, and
reenforces the assumption made in the fitting procedure that the
$\langle 0^{+}||E2||2^{+}\rangle$ matrix element can be coupled to
the $\langle 2^{+}||E2||4^{+}\rangle$ matrix element assuming the
validity of the rotational model.

\newpage

Table~\ref{results_E1} gives the values of $E$1 matrix elements for
$^{222}$Ra and $^{228}$Ra obtained in this work.

\begin{table}[H]
 \caption{\label{results_E1} Values of $E$1 matrix elements measured in the present experiment.
 The intrinsic moments, $\mathcal{Q}_1$, are derived from each matrix element (m.e.) using $\langle I_i || \mbox{$\cal{M}$} ($E$1) || I_f \rangle = \sqrt{(2I_i
+1)} \sqrt{3/4\pi} (I_i 0 1 0 | I_f 0) Q_{1} $. The signs of the
matrix elements are given by the rotational model, on the assumption
that $\mathcal{Q}_1$ has the same sign as $\mathcal{Q}_3$ for
$^{222}$Ra and the opposite sign for $^{228}$Ra, see the text of the
main paper.
 The uncertainties include the $1\sigma$  statistical error from the fit ($\chi^2 + 1$ type) and the systematic contributions. }
 \begin{ruledtabular}
\begin{tabular}{ccccc}

& $^{222}$Ra & & $^{228}$Ra & \\
$\langle I ||E1|| I'\rangle$ & m.e. & $\mathcal{Q}_1$ & m.e.   & $\mathcal{Q}_1$ \\
&  (eb$^{1/2}$) &  (efm) & (eb$^{1/2}$) &  (efm) \\
\hline
$\langle 0^{+}||E1||1^{-}\rangle$ & $0.0082 \pm 0.0007$ & $0.168 \pm 0.014$ & $-0.0043^{+0.0020}_{-0.0032}$ & $-0.09 ^{+0.04}_{-0.07}$\\
$\langle 2^{+}||E1||1^{-}\rangle$ & $-0.0114^{+0.0009}_{-0.0025}$   & $0.165 ^{+0.037}_{-0.013}$ & $0.0051^{+0.0033}_{-0.0022}$ & $-0.07^{+0.03}_{-0.05}$\\
$\langle 2^{+}||E1||3^{-}\rangle$ & $0.026 \pm 0.005$  & $0.31 \pm 0.06$ & $-0.009 \pm 0.004$ & $-0.11 \pm 0.04$ \\
$\langle 4^{+}||E1||3^{-}\rangle$ & $-0.027 \pm 0.026$  & $0.28 \pm 0.26$ & $0.0080^{+0.0036}_{-0.0023}$ & $-0.08^{+0.02}_{-0.04} $ \\
$\langle 4^{+}||E1||5^{-}\rangle$ & $0.0180^{+0.0048}_{-0.0019}$ & $0.165^{+0.044}_{-0.018} $ & $-0.013 \pm 0.006$ & $-0.12 \pm 0.05$ \\
$\langle 5^{-}||E1||6^{+}\rangle$ & $0.0277 \pm 0.0011$ & $0.231 \pm 0.010 $ & $-0.0074_{-0.0018}^{+0.0012}$ & $-0.061^{+0.010}_{-0.015}$ \\
$\langle 6^{+}||E1||7^{-}\rangle$ & $0.0273 \pm 0.0021$ & $0.211 \pm 0.016$ &  &   \\
$\langle 7^{-}||E1||8^{+}\rangle$ & $0.0317 \pm 0.0018$  & $0.230 \pm 0.013$ & $-0.013_{-0.010}^{+0.033}$ & $-0.09 ^{+0.24}_{-0.08}$ \\
$\langle 8^{+}||E1||9^{-}\rangle$ & $0.048 \pm  0.004$  & $0.326 \pm 0.025$& $-0.020^{+0.009}_{-0.020}$ & $-0.14 ^{+0.06}_{-0.14}$\\
$\langle 9^{-}||E1||10^{+}\rangle$ & $0.040 \pm 0.005$  & $0.26 \pm 0.03$ &  & \\
$\langle 10^{+}||E1|11^{-}\rangle$ & $0.050 \pm  0.006$   & $0.31 \pm 0.04$ &  & \\
$\langle 11^{-}||E1||12^{+}\rangle$ &  $0.031 \pm 0.004$  & $0.182 \pm 0.022$ &  & \\

 \end{tabular}

 \end{ruledtabular}
 \end{table}


\end{document}